\documentclass[journal, 10pt]{IEEEtran}
\IEEEoverridecommandlockouts
\usepackage{cite}
\usepackage{amssymb,amsfonts}
\usepackage{graphicx}
\usepackage{textcomp}

\usepackage{booktabs} 
\usepackage{amsthm}
\usepackage{amsmath}
\usepackage{array}
\usepackage{multirow}
\usepackage{epstopdf}
\usepackage{tcolorbox}
\usepackage{color, ulem}
\usepackage{colortbl}
\usepackage{adjustbox}
\definecolor{royalazure}{rgb}{0.0, 0.22, 0.66}
\definecolor{cardinal}{rgb}{0.77, 0.12, 0.23}
\usepackage{enumitem}
\theoremstyle{definition}

\usepackage{caption}

\usepackage{etoolbox}
\usepackage{tikz}
\usetikzlibrary{tikzmark}
\usetikzlibrary{calc}
\setlist[itemize]{leftmargin=*}

\usepackage{listings}
\usepackage{xcolor}

\definecolor{codegreen}{rgb}{0,0.6,0}
\definecolor{codegray}{rgb}{0.5,0.5,0.5}
\definecolor{codepurple}{rgb}{0.58,0,0.82}
\definecolor{backcolour}{rgb}{0.95,0.95,0.92}

\lstdefinestyle{mystyle}{
    backgroundcolor=\color{backcolour},   
    commentstyle=\color{codegreen},
    keywordstyle=\color{magenta},
    numberstyle=\tiny\color{codegray},
    stringstyle=\color{codepurple},
    basicstyle=\ttfamily\footnotesize,
    breakatwhitespace=false,         
    breaklines=true,                 
    captionpos=b,                    
    keepspaces=true,                 
    numbers=left,                    
    numbersep=5pt,                  
    showspaces=false,                
    showstringspaces=false,
    showtabs=false,                  
    tabsize=2
}

\newtcolorbox{boxA}{
    fontupper = \bf,
    boxrule = 1.5pt,
    colframe = black 
}

\newtcolorbox{boxB}{
    fontupper = \bf\color{main}, 
    boxrule = 1.5pt,
    colframe = main,
    rounded corners,
    arc = 5pt   
}

\newtcolorbox{boxC}{
    colback = sub, 
    boxrule = 0pt  
}

\newtcolorbox{boxD}{
    colback = sub, 
    colframe = main, 
    boxrule = 0pt, 
    toprule = 3pt, 
    bottomrule = 3pt 
}

\newtcolorbox{boxE}{
    enhanced, 
    boxrule = 0pt, 
    borderline = {0.75pt}{0pt}{main}, 
    borderline = {0.75pt}{2pt}{sub} 
}

\newtcolorbox{boxF}{
    colback = sub,
    enhanced,
    boxrule = 1.5pt, 
    colframe = white, 
    borderline = {1.5pt}{0pt}{main, dashed} 
}

\newtcolorbox{boxG}{
    enhanced,
    boxrule = 0pt,
    colback = sub,
    borderline west = {1pt}{0pt}{main}, 
    borderline west = {0.75pt}{2pt}{main}, 
    borderline east = {1pt}{0pt}{main}, 
    borderline east = {0.75pt}{2pt}{main}
}

\newtcolorbox{boxH}{
    colback = sub, 
    colframe = main, 
    boxrule = 0pt, 
    leftrule = 6pt 
}
\usepackage{comment}

\usepackage[linesnumbered,ruled,vlined]{algorithm2e}
\usepackage[noend]{algpseudocode}

\def\BibTeX{{\rm B\kern-.05em{\sc i\kern-.025em b}\kern-.08em
    T\kern-.1667em\lower.7ex\hbox{E}\kern-.125emX}}
\begin{document}

\title{JARVIS: A Multi-Agent Code Assistant for High-Quality EDA Script Generation}

 \author{
 \IEEEauthorblockN{Ghasem Pasandi, Kishor Kunal, Varun Tej, Kunjal Shah, Hanfei Sun, Sumit Jain, Chunhui Li, Chenhui Deng, Teodor-Dumitru Ene, Haoxing Ren, and Sreedhar Pratty} \\
 \IEEEauthorblockA{
 \textit{Nvidia Corp.}\\
 Santa Clara, CA, USA\\
 \{gpasandi, kkunal, vtej, kunjals, hanfeis, sumjain, chunhuil, cdeng, tene, haoxingr, spratty\}@nvidia.com}
 
 }

\maketitle

\begin{abstract}
This paper presents JARVIS, a novel multi-agent framework that leverages Large Language Models (LLMs) and domain expertise to generate high-quality scripts for specialized Electronic Design Automation (EDA) tasks. By combining a domain-specific LLM trained with synthetically generated data, a custom compiler for structural verification, rule enforcement, code fixing capabilities, and advanced retrieval mechanisms, our approach achieves significant improvements over state-of-the-art domain-specific models. Our framework addresses the challenges of data scarcity and hallucination errors in LLMs, demonstrating the potential of LLMs in specialized engineering domains. We evaluate our framework on multiple benchmarks and show that it outperforms existing models in terms of accuracy and reliability. Our work sets a new precedent for the application of LLMs in EDA and paves the way for future innovations in this field.
\end{abstract}

\section{Introduction}

Large Language Models (LLMs) have revolutionized software development by automating various coding tasks, streamlining repetitive processes, and enhancing developer productivity. Tools like Microsoft's Copilot and Meta's CodeLlama have demonstrated the potential of LLMs in generating boilerplate code, automating common patterns, and embedding best practices within generated outputs \cite{Codex, code_understanding, Qin2023ToolLLM}. However, when applied to specialized fields like Very-Large-Scale Integration (VLSI) design within Electronic Design Automation (EDA), LLM performance is hindered by the scarcity of relevant training data, leading to unreliable and inaccurate outputs. These models often misinterpret and hallucinate due to a lack of contextual depth, highlighting the need for domain-specific fine-tuning.

Recent efforts have focused on enhancing the reasoning capabilities of LLM models using CoT \cite{wei2023CoT} and agent-based frameworks \cite{yao2023react} for general tasks. In the context of EDA, TOOLLLM \cite{Qin2023ToolLLM} uses a fine-tuned LlaMa model and integrates a depth-first search-based decision tree (DFSDT) to bolster planning and reasoning abilities. Additionally, Retrieval-Augmented Generation (RAG) techniques have been explored in recent works\cite{Ansys2024EngCode, jain2023CodeCleaning, RAG_based_codegen, jain2024llmagents}, which demonstrated the utility of retrieval-augmented control code generation in improving LLM performance for niche software engineering tasks. However, these approaches have limitations when applied to highly specialized domains like EDA, where complex dependencies and domain-specific knowledge are crucial.

To address the unique challenges of EDA script generation, we propose a novel framework that focuses on four key areas: Domain Adapted Pre-Training (DAPT) and Domain Supervised Fine-Tuning (DSFT) for model training, multi-LLM collaboration, custom compiler development, and multi-agent feedback flow. Our approach leverages a custom compiler that converts the custom VLSI tool manual to a knowledge graph using Abstract Syntax Tree (AST) format and provides augmented feedback data to the coding agent. This is complemented by a Synthetic Data Generation (SDG) approach to further improve access to domain-specific knowledge. SDG has been shown to significantly boost LLM performance in various applications \cite{guo2024SDG}. Our framework also utilizes RAG models, which effectively integrate domain-specific knowledge into the generation process by retrieving relevant contextual data from external sources.
\begin{figure}[t]
    \centering
    \includegraphics[width=3.3in]{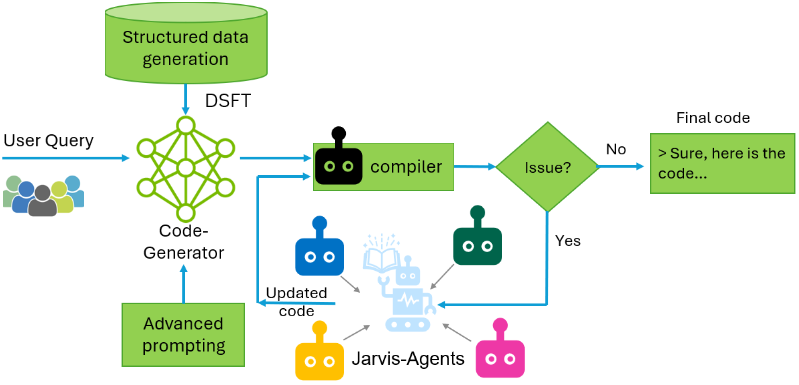}
    \caption{Overview of the JARVIS framework, illustrating different components and the feedback loop, enabling iterative code refinement and improvement.}
    \label{fig:overview}
\end{figure}

We developed a multi-LLM framework that synergistically leverages the strengths of diverse LLMs to generate high-quality EDA scripts. Our approach includes a custom compiler that detects and diagnoses issues in EDA code, providing actionable feedback for correction and streamlining the debugging process, with a focus on mitigating hallucination errors. We also introduce a multi-episode, multi-agent framework using ReAct, which optimizes initial EDA scripts using multiple custom-developed tools and a feedback loop, demonstrating the potential of reactive and multi-agent approaches in EDA code improvement.

Our framework, called JARVIS (Just A Remarkable VLSI Intelligence System), demonstrates substantial improvements over existing models for EDA script generation. The proposed framework has broader implications, setting a precedent for the future application of LLMs in other specialized engineering domains. We evaluated the capabilities of various LLMs, including ChatGPT with GPT4o, and LlaMa 3.1.

Fig. \ref{fig:overview} illustrates the components of our framework. The framework consists of a code generator which uses an LLM model trained on domain specific data to generate initial codes for user queries, then through collaboration of multiple jarvis-agents, and custom tools like code compiler, the code is being refined by mitigating hallucination errors and producing high-quality EDA scripts.

The main contributions of this work are as follows:

\begin{itemize}
    \item A novel Supervised Fine-Tuning (SFT) pipeline for LLMs, enhancing their performance in data-scarce environments and unlocking their potential for EDA tasks.
    \item A multi-LLM framework that collaboratively leverages diverse LLMs to generate high-quality EDA scripts.
    \item A custom code compiler that detects and diagnoses issues in an EDA code, providing actionable feedback for correction and streamlining the debugging process.
    \item A multi-agent multi-episode framework using ReAct, which optimizes initial EDA scripts using multiple custom-developed tools and a feedback loop.
    \item Comprehensive experimental results on multiple LLMs, validating the effectiveness of our approach on real-world VLSI/EDA tasks.
\end{itemize}

\section{Related Work}
The capabilities of LLMs in coding related tasks such as test-driven code generation \cite{fakhoury2024ltestdriven}, code refactoring \cite{jain2023CodeCleaning, long2024SDG}, and generating synthetic data for training purposes \cite{long2024SDG} have been well-documented. 
To enhance the quality of LLM-generated code, research is being conducted in several directions. These include model improvement using various training techniques and Self-Distillation with Guidance (SDG), leveraging retrieval to enhance domain knowledge, improving reasoning abilities through prompt-tuning, and implementing corrective-feedback flows using tools or multi-LLM-based evaluation.

New LLM models targeting coding applications such as CodeLlama\cite{CodeLlama}, StarCoder\cite{StarCoder}, StableCode\cite{StableCode} are being trained to improve LLMs code generation capabilities. 
SDG approaches \cite{long2024SDG, liu2024chatqa, guo2024SDG, golang} have become popular in enhancing LLM’s capability of integrating user-provided or retrieved context for conversational tasks.
RAG techniques have been explored in recent works\cite{Ansys2024EngCode, jain2023CodeCleaning, RAG_based_codegen, jain2024llmagents}, which demonstrated the utility of retrieval-augmented control code generation in improving LLM performance for niche software engineering tasks. To further reduce hallucinations in code generation, approaches like AST structure-aware pretraining \cite{gong2024astt5} boost the performance of coding tasks.

Recent efforts have focused on enhancing the reasoning capabilities of LLM models using Chain-of-Thought (CoT) \cite{wei2023CoT}, ReAct framework \cite{lin2024SoftwareUse, fakhoury2024ltestdriven} and agent-based frameworks \cite{yao2023react} for general tasks. To improve reasoning, LLM agents can leverage external models, tools, plugins, or APIs to tackle complex problems \cite{NEURIPS2023_AGI}. 
The integration of feedback loops and agent-based frameworks has been pivotal in improving LLM performance for real-world tasks, especially in niche domains. Works like CodeAgent\cite{zhang2024codeagent, execution} demonstrate the effectiveness of feedback-driven mechanisms in refining LLM outputs through execution-based verification and active user interaction.

In the EDA domain, creating scripts for custom VLSI tools presents unique challenges. The complex, domain-specific requirements of EDA, combined with a lack of representative training data, limit the effectiveness of general-purpose LLMs like Copilot and CodeLlama. Recent advancements in domain-adapted models, such as ChipNeMo\cite{chipnemo, Qin2023ToolLLM}, illustrate the potential of LLMs tailored to chip design tasks. ChipNeMo employs several domain adaptation techniques, including domain-adaptive tokenization, domain-adaptive pretraining (DAPT), and model alignment, to customize LLMs for chip design. TOOLLLM\cite{Qin2023ToolLLM} uses a fine-tuned LlaMa model and integrates a depth-first search-based decision tree to enhance planning and reasoning abilities.

However, ChipNeMo's performance can be further improved through additional specialized fine-tuning (SFT) using data generated by methods like SDG. TOOLLLM's performance can also be enhanced using an agentic framework, which provides better reasoning capabilities. These innovations underscore the importance of domain-specific tasks, such as EDA, where data scarcity remains a persistent issue. Despite these efforts, significant gaps remain in applying these models to highly specialized domains like EDA. In this paper, we address some of these challenges by exploring advanced domain adaptation techniques and proposing new methodologies to enhance the performance of LLMs in the EDA domain.

\section{Synthetic Data Generation (SDG)}
\label{sec:SDG}
DAPT and DSFT are essential components of model training. DAPT uses raw data during training and DSFT uses a custom-labeled data for fine-tuning a model. The SFT data should encompass deep knowledge of a specific domain, enabling the LLM to grasp the nuances necessary for handling specialized tasks. This requires high quality labeled data. We experimented with a small set of manually generated high quality data but we observed signs of overfitting in the trained model due to the limited domain SFT data.

Specialized SDG approaches can enhance the effectiveness of SDG by controlling the data chunking process or making small perturbations in existing codes \cite{long2024SDG}. However, due to the limited reasoning capabilities of LLMs, it is often unrealistic to expect them to generate an entire code dataset, especially when the data involves complex structures or semantics. Our observations indicate that while LLMs struggle with code generation tasks for custom tools, they excel at generating comments and summarizing tasks. Therefore, we leveraged LLMs to generate comments and questions from raw codes, thereby increasing our SFT data examples.

\begin{figure}[t]
    \centering
    \includegraphics[width=2.2in]{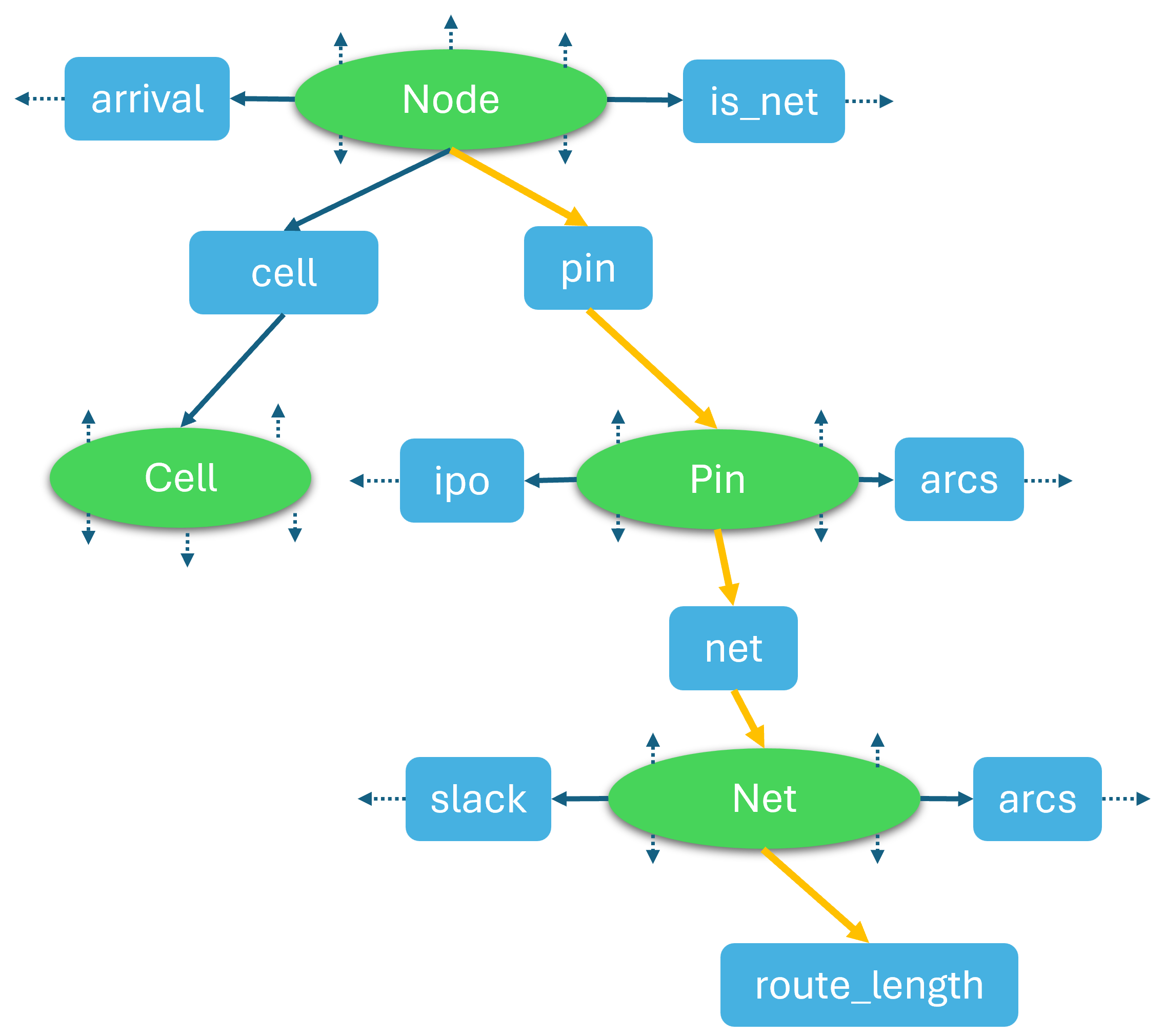}
    \caption{Example for a tool API graph constructed from man page.}
    \label{fig:graph}
\end{figure}

To overcome the data limitation and increase the tool command coverage, we developed a random code generator that utilizes the tool’s command graph as shown in an example in Fig. \ref{fig:graph} to create syntactically correct codes using an AST.
An AST is a tree representation of the abstract syntactic structure of source code written in a programming language. Each node of the tree denotes a construct occurring in the source code.
Thus, the high level idea of using an AST is to build random code stage by stage by selecting random objects, attributes and then unparse the structure to generate human-readable code. 
Listing \ref{code_comment_1} shows an example function. The parsing process converts it to an AST as shown in Listing \ref{code_comment_2} where each variable, operation, function calls is identified and stored in a graph structure. This AST structure makes the code-generation flow independent of any specific programming language, allowing it to be extended to other programming languages.   

\begin{lstlisting}[language=Python, numbers=none, label=code_comment_1, caption=An example function to demonstrate our random code generation process.]
def random_function():     
    pins_obj = get_all_pins()
    return pins_obj[5]
\end{lstlisting}

\begin{lstlisting}[language=Python, numbers=none, label=code_comment_2, caption=AST for the example function shown in Linsting \ref{code_comment_1}.]
Module: entire code
    FunctionDef: Function Body
    Assign: pins_obj=get_all_pins()
        Name: pins_obj
        Call: get_all_pins()
Return: return pins_obj[5]
    Subscript: pins_obj[5]
        Name: pins_obj
        Constant: 5
\end{lstlisting}

The code generation algorithm is detailed in Algorithm \ref{algo_sdg}. The process begins by randomly selecting an operation and a few start nodes (design objects). The algorithm then traverses the nearby connected nodes of the start nodes to gather various attributes of the objects. It applies the selected operation on an attribute with a valid data type matching the selected operation. The function parameters are then initialized with random values and design objects. After several iterations, the AST is converted into a code. By combining this approach with LLM-based comment and code generation, we created an additional 35,000 SFT training data points to cover most of the commands in our target EDA tool.

\begin{algorithm}[t]
{\small
\caption{Random Synthetic Code Generation Flow}
\begin{algorithmic}[1]
    \State Select Code structure by randomly choosing one operation (such as condition, math operations, iterators, OOPs operations)
    \State Randomly Select and Initialize Classes, Attributes, from tools API graph based on chosen operation
    \State Apply operation on selected class attributes and store in an AST format.
    \State Initialize function parameters and arguments.
    \State Repeat steps 2-5 for random iterations
    \State Convert AST to Code
    \State Add line by line comments using LLM
    \State Generate Questions based on commented code using LLM
\end{algorithmic}
\label{algo_sdg}
}
\vspace{-.25em}
\end{algorithm}

On these code snippets we extracted the APIs used along with their man page description and used the Nemotron 340B model \cite{nvidia2024nemotron4340b} to add comments to the code and generate questions.
Listing \ref{code_comment_3} shows an example of generated code along with question and comments.
\begin{lstlisting}[numbers=none]
Question: 
Write a code to find the largest logic delay among a set of violations.
\end{lstlisting}

\begin{lstlisting}[language=Python, numbers=none, label=code_comment_3, caption=Example of a randomly generated code that commented and a question is generated for it.]
# Get the set of violations
vios_obj_1 = get_violations('*')
# Initialize the largest logic delay to a 0
largest_logic_dly = 0
# Iterate over each violation in the set
for vio in vios_obj_1:
        # Compare the current value to largest delay
        if vio.logic_delay() > largest_logic_delay:
            # Update the largest delay
            largest_logic_dly = vio.logic_delay()
    # Print the largest logic delay
    print(largest_logic_dly)
\end{lstlisting}

Once the synthetic SFT dataset is generated, we finetune the ChipNeMo model on it to enhance its understanding of complex structures in VLSI tool scripts, employing the standard autoregressive language modeling objective. We set the learning rate to $5e$-$6$ using a constant scheduler, apply a weight decay of $0.1$, and a batch size of $128$. Additionally, we limit training to one epoch to prevent overfitting. The training process utilizes $16$ nodes in a compute cluster, each equipped with eight A100 GPUs boasting 80GB of memory. It completes in approximately eight hours. 
\section{Multi-Agent Framework}
\label{sec:agents}
\begin{figure}[t]
    \centering
    \includegraphics[width=3.5in]{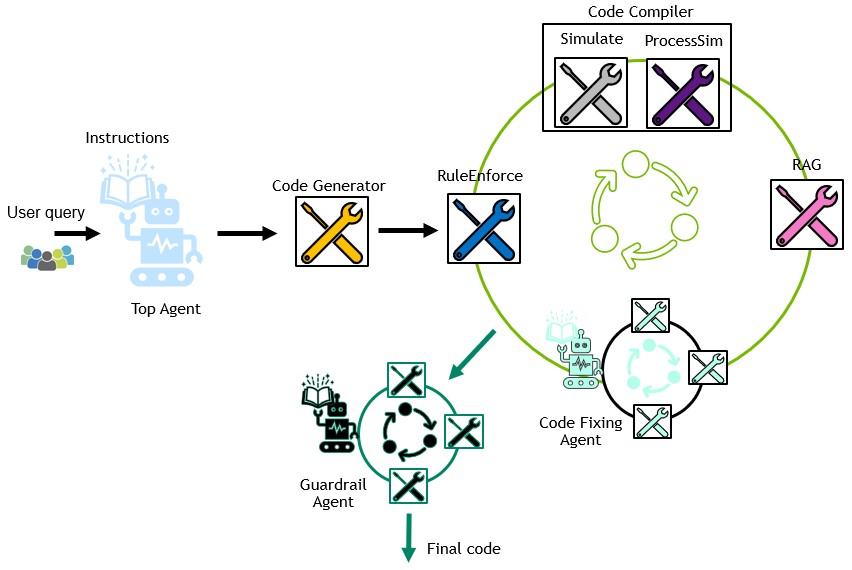}
    \caption{Overview of the multi-agent flow, listing the various tools utilized to improve code quality: Code Generator, RuleEnforce, Code Compiler, Code Fixing Agent, RAG, and a Guardrail agent.}
    \label{fig:overview_react}
\end{figure}

We present a novel multi-agent framework that leverages the strengths of state-of-the-art (SOTA) LLMs and domain-adapted models to generate high-quality scripts for custom VLSI tools. Our framework builds upon the ReAct framework \cite{yao2023react}, which we modified to utilize SOTA LLMs like LLaMa 3.1.
To further enhance the framework's capabilities, we developed a multi-agent multi-episode flow that integrates multiple custom tools. Our approach combines the generalized linguistic capabilities of LLMs with the specialized domain knowledge of domain-adapted models like ChipNeMo. As illustrated in Fig. \ref{fig:overview_react}, our multi-agent flow consists of a Top Agent, a Code Fixing Agent, a Guardrail Agent and a set of tools including Code Generator, RuleEnforce, Code Compiler, and RAG, to ensure the generation of high-quality code that meets output standards. In the following, we will provide more details on each of these components. The Top Agent is a ReAct agent with instructions on how to generate high quality EDA scripts and with access to the said set of tools and ability to communicate with other agents in the loop.

\subsubsection{Code Generator}
We use ChipNeMo as our Code Generator to generate the first answer for any user query. For other tasks, if an LLM is needed, we use a SOTA LLM like LlaMa 3.1 and not ChipNeMo as we observed challenges with conversation follow-up with domain-trained models as they had limited conversation data during training. This challenge can be attributed to missing Preference Fine-Tuning such as Direct Preference Optimization \cite{rafailov2024DPO}, and we plan to improve on this. This alignment process enables the model to follow instructions better and engage in conversations effectively.

\subsubsection{RuleEnforce}

Despite the advancements of DAPT and SFT, we observed that LLMs consistently struggle with certain topics, often resulting in partial or incorrect implementations. For example, they may not know how exactly to compute power in the target EDA tool. To address this limitation, we developed RuleEnforce, a tool in the Top Agent's disposal that leverages special rules, hints, and notes of the target EDA tool. By defining explicit rules and instructions, RuleEnforce enables the agent to accurately apply these rules to relevant code segments.

We employ two methods for rule extraction: manual and auto rule extraction. In the manual approach, we identify common patterns of errors in codes written by the flow through extensive experiments. For instance, we observed that power consumption calculations often require special handling, as illustrated in Listing \ref{rule_enforce_example}. We wrote a set of rules manually to address these issues, such as the power computation rules shown in Listing \ref{power_comp_rules}.
In contrast, our auto rule extraction method utilizes a ReAct agent that analyzes a set of QnA data with user queries and golden codes. The agent extracts a set of rules that can generate codes similar to the golden codes, while also consulting with our simulator to prevent simulation errors. This process is done offline and yields thousands of rules, which are then filtered and ranked by a RAG system using the user query during the run time. The top 10 related rules are appended to our manual rules, enhancing the overall accuracy of the RuleEnforce tool.

Listing \ref{rule_enforce_example} demonstrates the effectiveness of RuleEnforce in correcting code errors. The upper code block contains an issue in calculating power (line 4), as the target EDA tool does not have an attribute named leakage\_power for cell. After applying the special rules for computing power, as shown in Listing \ref{power_comp_rules}, the code is revised to produce the corrected output, as demonstrated in the lower code block.
Section \ref{sec:evaluator} will explain how our Code Compiler can help fix a hallucinated attribute. However, in this example, using only the Code Compiler will not resolve the issue. This is because, although the compiler can detect leakage\_power as an invalid attribute for cell and suggest related attributes such as power, knowing valid attributes alone is not sufficient. The power computation requires an additional step, as demonstrated above.

\begin{lstlisting}[language=Python, numbers=none, caption=]
total_leakage_power = 0
for cell in get_cells("*", "hierarchical"):
    if cell.is_sequential():
        leakage_power = cell.leakage_power
        total_leakage_power += leakage_power
\end{lstlisting}

\begin{lstlisting}[language=Python, numbers=none, label=rule_enforce_example, caption=Example for before and after applying power computation special rules.]
total_leakage_power = 0
for cell in get_cells("*", "hierarchical"):
    if cell.is_sequential():
        cell.calculate_power()
        leakage_power = cell.power("is_leakage")
        total_leakage_power += leakage_power
\end{lstlisting}

\begin{lstlisting}[language=Python, numbers=none, label=power_comp_rules, caption=Special rules for computing cell power.]
You can compute power of a cell as below:
    + First use Cell.calculate_power() to compute power values.
    + Next, access desired power type using flag with the following values:
        - power.is_leakage
        - power.is_dynamic
        - power.is_total
        - power.is_switching
    + Examples:
        - cell.calculate_power()
          leakage = cell.power("is_leakage")
        - cell.calculate_power()
          switching = cell.power("is_switching")
\end{lstlisting}

\subsubsection{Code Compiler}
\label{sec:evaluator}

LLMs for code generation tasks are prone to mistakes such as incorporating wrong logic, inefficient code, misunderstanding the full intent of the question, and hallucinating. Addressing hallucinations and object-attribute relationship-related issues in limited data scenarios proved challenging because the model does not understand the relationship between various objects and attributes. To fix these issues, we developed an AST-based custom compiler to process LLM-generated code and provide corrective feedback to the model.

We have been focused on an internal high-usage EDA tool, which is object-oriented, interacts in Python, and has many attributes for each object. We use the tool's man page to generate a tool command graph similar to what is shown in Fig. \ref{fig:graph}. The green nodes represent the data-objects, and blue boxes represent the attributes available on the object. We can use these attributes to traverse across different objects, e.g., we can use `Pin.net()` to get the `net` object connected to a `pin`.

Our compiler has two main jobs: (1) simulating the code and (2) processing the simulation results and providing useful feedback. For the simulation part, it converts the given initial code into an API-graph and walks through each node. While walking through the nodes, it checks for the existence of each object and attribute in the API-graph and checks for the compatibility of this object-attribute relationship. It returns any lines of the code where there is incompatibility or usage of hallucinated APIs. This is done using the `Simulate` tool in Fig. \ref{fig:overview_react}.
Next, in case of hallucination, our compiler returns the list of valid attributes for the corresponding object, and in case of wrong object-attribute relationship, it returns the shortest path from the object to the attribute. This is done using the `ProcessSim` shown in Fig. \ref{fig:overview_react}.

Listing \ref{sim_example_1} shows an example for our compiler. It shows a user query, an initial code for it, simulation results, and compiled data for it, followed by refined code. As seen, the shortest path provided by the compiler is used to fix the erroneous line and make it error-free.
\begin{lstlisting}[numbers=none]
User query: 
Write a code to get all hold violations, if any net in the vio has a route length greater than 2um.
\end{lstlisting}

\begin{lstlisting}[language=Python, numbers=none, caption=]
for node in nodes:
    if not(node.is_net()):
        if node.route_length() > 2:
            filtered_hold_vios.append(vio)
\end{lstlisting}

\begin{lstlisting}[language=Python, numbers=none, caption=]
# Simulation results:
Line No. 3: node of datatype Node has no attribute route_length
# Valid attributes and shortest paths:
Here are a few valid attributes on Node:
  Node.pin => Netlist pin object
  Node.pin_name => Pin name 
  Node.pin_report => Report pin object
Here is how to get to route_length from Node:
  Node -> pin -> Pin -> net -> Net -> route_length
\end{lstlisting}

\begin{lstlisting}[language=Python, numbers=none, label=sim_example_1, caption=Example showing how our compiler works and what data it provides.]
for node in nodes:
    if not(node.is_net()):
        if node.pin().net().route_length() > 2:
            filtered_hold_paths.append(path)
\end{lstlisting}

\subsubsection{RAG for domain data}

To enhance our script generation framework, we incorporate RAG, leveraging an open-book strategy to retrieve in-domain knowledge from external data stores, enabling the LLM to generate more precise and contextually grounded responses. A common use of RAG is augmenting the initial query with retrieved code examples. However, we observed that providing full code snippets can lead to overfitting in cases where there is a significant gap between the user query and retrieved code. To mitigate this, we optimize RAG for API queries from tool documentation and single-line code retrieval with high accuracy.

Our implementation employs a hybrid search approach, combining BM25 for precise keyword matching with FAISS for capturing semantic relationships. To further improve accuracy, we integrate a re-ranker that merges results from both models and prioritizes the most relevant matches. This RAG framework enhances LLM performance while reducing hallucinations by providing more contextually relevant information.

\subsubsection{Code Fixing Agent}

It often happens that we have a script that is mostly correct but there are one or few issues in some lines. These issues sometimes arise because the model had good ideas of what needed to be done but couldn't do it in a quite accurate way. For example, it was trying to get the top reference of a design, but didn't use the right APIs or made a mistake on its correct usage. If we knew what was the purpose of an erroneous line, then a small query that specifically asks for that can fetch a small code snippet from our domain expert LLM, ChipNeMo-jarvis, and pass it to the ReAct agent to be used in next iterations for fixing the issues. The intent of a line can be understood systematically by looking up its surrounding comments and analyzing those effectively. What is described is exactly what the code fixing agent does in our framework. We implemented it as a tiny ReAct agent with access to ChipNeMo-jarvis as a tool. This way of implementation was intended to make the code fixing calls more effective and efficient.

\subsubsection{Guardrail Agent}

To ensure the quality of the generated code, we introduced a `guardrail agent` that scrutinizes the output before presenting it to the user. This agent is integrated with a LlaMa 3.1 LLM and evaluates the code's structural and functional correctness. Structurally, our compiler assesses the accuracy of API calls, assigning a score based on the percentage of correct API calls and object-attribute relationships. Functionally, the agent converts the final code's comments into a question and compares it to the user's original query, assigning a score accordingly. Only if the code meets both structural and functional quality checks, it is presented to the user; otherwise, a standardized message, such as `\textit{Sorry, I don't have enough knowledge to write such a code for you}` is displayed, indicating that the tool is unable to provide a satisfactory response.

\begin{lstlisting}[numbers=none]
User query: 
Write a code to find the violation ID of a path with worst negative slack among vios ending in RAM
\end{lstlisting}

\begin{lstlisting}[language=Python, numbers=none, floatplacement=t, caption=]
def get_worst_slack_of_viol(vpaths):
    w_slack = vpaths.sort_using_slack()[0].slack()
    for v in vpaths:
        if v.is_setup_vio():
            slack = v.slack()
            if slack < w_slack:
                worst_slack = slack
            end_pin = v.get_end_pin()
            if end_pin.of_ram():
                return w_slack
    return w_slack
# Call the function with the given vio_id
worst_slack = get_worst_slack_of_viol(get_all_violations())
\end{lstlisting}

\begin{lstlisting}[language=Python, numbers=none, floatplacement=t, label=complex_example, caption=A complex example showcasing how our flow improves initial code by using multiple custom tools]
# Get all vios
vios = get_all_violations()
# Filter vios that end at RAM
ram_vios = [v for v in vios if vio.get_end_ref().of_ram()]
ram_vios_v = Violations()
for vio in ram_vios:
    ram_vios_v.push_back(vio)
# Sort vios by slack and get the worst one
worst_slack_vio = ram_vios_v.sort_using_slack()[0]
worst_slack_vio_id = worst_slack_vio.id()
\end{lstlisting}

Listing \ref{complex_example} shows a relatively complex example on how our flow improves the quality of the code by leveraging multiple custom agents. In this example, the user wants to find a VIO with the worst negative slack among all VIOs ending at RAM. The initial code has incorrect object-attribute relationships, `get\_end\_pin` and `of\_ram` in line 9 of the upper code block, and wrong logic, returning the first VIO ending in RAM instead of the vio with the worst slack.
Our flow fixes this by using multiple tools. Our compiler catches the incorrect object-attribute relationship and provides valid attributes. Our code fixing agent catches the intent of the erroneous line, accessing the end reference object of a VIO, and fetches a piece of code for that. Finally, our RuleEnforce tool realizes that we want to sort VIOs and get the worst one; it uses a provided hint to replace the for loop with a single line of code to do the same, which makes the code efficient.

The pseudo code for our multi-agent based code refinement algorithm is presented in Algorithm \ref{algo_vivid_python}. This algorithm, as detailed above, formalizes the iterative refinement process, which involves formatting, rule enforcement, and simulation, as well as the integration of multiple agents to resolve errors and improve code quality. 
We implemented a multi-episode version of the discussed algorithm, where Algorithm \ref{algo_vivid_python} is called in each episode. This approach effectively addresses several issues present in the native ReAct implementation from Lanchain, including the tendency to trap into infinite or repeated loops where the agent becomes stuck calling the same tool without a clear termination condition. By defining each episode as a sequence of multiple rounds of tool calls and iterative refinements of the code, we obtain a summary of the overall progress at the end of each episode which is fed back in at the beginning of the next episode. This multi-episode implementation has improved the reliability and efficiency of the flow.

\begin{algorithm}[t]
{\small
\caption{Multi-Agent based Code Refinement Algorithm}\label{algo_vivid_python}
\DontPrintSemicolon 
\KwIn{Query: a coding query by user, TimeLimit, ItrLimit} 
\KwOut{Refined code}
Initialize Top Agent \\
$InitCode$ = Code\_Generator($Query$) \\
$RefinedCode$ = $InitCode$ \\
$SimResult$ = Not Clean, $TimeElapsed$ = 0, $ItrCount$ = 0 \\
\While{$SimResult$ != Clean and $TimeElapsed <= $ TimeLimit and $ItrCount <= $ ItrLimit}{
    $RefinedCode$ = RuleEnforce($RefinedCode$) \\
    $SimResult$ = Simulate($RefinedCode$) \\
    \If{$SimResult$ != Clean}{
        $NewCodes$ = \{\} \\
        \For{each erroneous line in $SimResult$}{
            $NewCode$ = Code\_Fixing\_Agent($RefinedCode$, erroneous line) \\
            Append($NewCodes$, $NewCode$) \\
        }
        $ValidAttrs$, $ShortPaths$ = ProcessSim($SimResults$)
    }
    $RefinedCode$ = Top\_Agent($NewCodes$, $SimResult$, $ValidAttrs$, $ShortPaths$) \\
    $TimeElapsed$ = $TimeElapsed$ + $t$ \\
    $ItrCount$ = $ItrCount$ + 1 \\
}
$Output$ = Guardrail\_Agent($RefinedCode$) \\
\Return{$Output$}
}
\vspace{-.25em}
\end{algorithm}
\section{Experimental Results}
We evaluated our method on multiple benchmarks: B-easy (simple API query) with 150 questions, B-medium (2-3 stages of reasoning) with 30 questions, and B-hard (multiple reasoning stages) with 20 questions. The results are summarized in Tables \ref{tab:final_results}-\ref{tab:ablation}.

To generate the answers in our evaluation, we set the temperature to zero which allowed for more predictable and deterministic code generation, which is essential for tasks requiring high accuracy. We used pass@1 accuracy for our evaluations. B-easy was evaluated using an auto-evaluation method using string match for the desired APIs, while B-medium and B-hard required human evaluation due to the dependency of code executions on design data.

As we can see, GPT4o and LlaMa 3.1 (with retrieval) achieved close to 0\% accuracy on all benchmarks, indicating that off-the-shelf models do not perform well on domain-specific data. This is because these models lack the specialized domain knowledge required for code generation tasks. There are a few instances where these models managed to generate correct answers based on their general VLSI knowledge and their probabilistic nature. However, these instances are rare and do not compensate for their overall poor performance in this domain.

\subsection{Domain training}
Domain adaptive training as presented in \cite{chipnemo} shows the importance of DAPT and DSFT for models to incorporate the domain information. The initial ChipNeMo trained model utilized existing raw data for DAPT and a manually created dataset of 150 question-and-answer pairs for DSFT. This model demonstrated a significant improvement of approximately 50\% over off-the-shelf models. To address the challenge of limited user data, we generated 35,000 synthetic data points for DSFT using small code snippets, as detailed in Section \ref{sec:SDG}. This synthetic data notably enhanced the model's accuracy, particularly on the easy benchmark. While there was a 7\% improvement on hard benchmarks, the most substantial gain was observed on the easy benchmark, with a 21\% increase. This improvement is attributed to the quality of the synthetic data, as small snippets were selected to avoid the incoherence and multiple local inherited functions that larger chunks tend to produce. 
\begin{table}[t]
\centering
\scriptsize
\caption{Evaluation results using different LLMs.}
\label{tab:final_results}
\begin{tabular}{|c|c|c|c|c|}
\hline
Model name	& B-easy &	B-medium & B-hard\\
\hline \hline
GPT4o  (with retrieval)	&0\%	& 0\%	& 10\%	\\
LlaMa 3.1 (with retrieval)	&4\%	& 0\%	& 7\%\\
DAPT+DSFT with manual data (ChipNeMo)	&46\%	& 56\%	& 36\%	\\
DAPT + DSFT with added synthetic data &67\%	& 62\%	& 43\%	\\
DSFT with added synthetic data &68\%	& 36\%	& 18\%\\
\hline
\end{tabular}
\vspace{-1.0em}
\end{table}

\begin{table}[t]
\centering
\scriptsize
\caption{Ablation studies for different agent components}
\label{tab:ablation}
\begin{tabular}{|c|c|c|c|c|}
\hline
Model name	& B-easy &	B-medium & B-hard\\
\hline \hline
\textbf{Full feature Multi-agent flow} & 92\% & 93\%	& 81\%\\
\midrule
LlaMa 3.1 with multi-agent flow & 53\% & 43\%	& 30\%\\
Multi-agent flow w/o RuleEnforce & 89\% & 84\%	& 56\%\\
Multi-agent flow w/o RAG & 73\%& 79\%	& 80\%\\
Multi-agent flow w/o code fixing agent & 90\% & 93\%	& 75\%\\
\hline
\end{tabular}
\vspace{-1.0em}
\end{table}
 
In another experiment to observe the impact of DSFT vs DAPT+DSFT, we conducted SFT training on the foundation model using the synthetic data. The DSFT only model provided similar gain on the easy benchmark (68\% with DAPT only vs 67\% with DAPT+DSFT), while the hard benchmark did not see any significant benefit. These results indicate that DSFT effectively helps the model learn domain-specific keywords. However, for complex reasoning tasks, DAPT remains crucial.
\begin{figure}[t]
    \centering
    \includegraphics[width=3.0in]{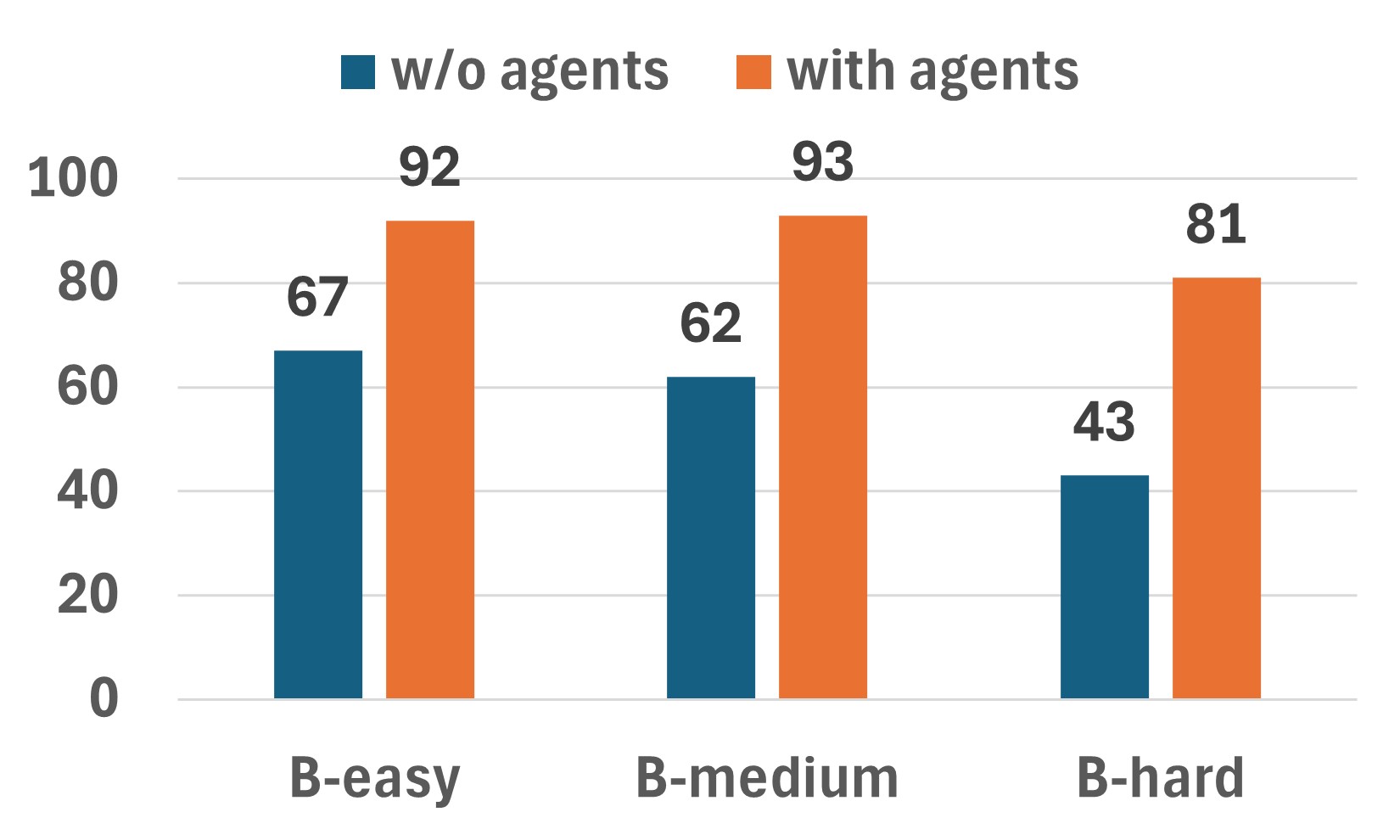}
    \caption{Comparing best scores without multi-agent flow ($4^{th}$ row in Table \ref{tab:final_results}) with the full featured multi agent flow. As seen significant improvements across all evaluated benchmarks are achieved.}
    \label{fig:exp_results}
\end{figure}
\subsection{Multi-agent component ablation studies}
We evaluated the effectiveness of the full featured multi-agent flow. The results are presented in Table \ref{tab:ablation}, the second row. Comparing it with Table \ref{tab:final_results}, the full featured multi-agent flow provides significant improvements on the quality of results across different evaluated benchmarks, as seen in Fig. \ref{fig:exp_results}. This demonstrates the great potential of our multi-agent flow.
We evaluated different agents and their components to understand their impact on performance. In the following, we provide ablation study on main components of JARVIS. 

\textbf{\uline{Initial ChipNeMo calls:}} For the feedback loop using agentic flow to work effectively, a good starting point is essential. The initial ChipNeMo call provides this foundation. As shown in Table \ref{tab:ablation}, starting with an off-the-shelf model like LlaMa 3.1 results in a slight improvement in agentic-feedback flow compared to the initial zero accuracy of LlaMa 3.1 in Table \ref{tab:final_results}. This improvement is due to the addition of more domain knowledge. However, the overall accuracy remains significantly lower compared to starting with the domain-trained model. 

\textbf{\uline{RuleEnforce:}} Codes generated for B-hard questions were often incorrect due to numerous internal rules to cross-reference data from different sources such as library, timing-models, physical layout information. These rules needs to be applied while writing complex codes. Easy and medium complexity questions have lesser dependency on these rules. Using RuleEnforce helped improve the B-hard accuracy by 25\% as seen in Table \ref{tab:ablation}.

\textbf{\uline{RAG for domain data:}} We employ retrieval to access data from tool documentation and existing code. While retrieval significantly improves performance on easy (19\%)  and medium (14\%) benchmarks, its impact on hard benchmarks is minimal (1\%). This highlights the need for high-quality, relevant data to reduce hallucinations. Complex reasoning remains challenging with simple documentation retrieval, as it requires understanding interactions between multiple components—especially keywords that may not be explicitly present in the user query.

\textbf{\uline{Code Fixing Agent:}} For complex codes, it would be better to provide an example code as a feedback along with the error information. So, we used our code fixing agent to fetch example code snippets for erroneous parts of initial codes. We have seen an improvement of about 6\% on B-hard by this process, while not much change in B-easy and B-medium. 

\section{Conclusions}
We formalize the high-quality EDA script generation task for data-scarce custom tools. To enable LLMs to tackle EDA script generation, we propose an innovative multi-LLM framework that integrates domain knowledge through specialized tools. This framework incorporates multiple custom tools such as custom code-compiler to refine initial obtained codes. Agent-based strategies are developed to optimize the use of these tools and achieve high quality final codes. 
To evaluate the effectiveness of our method, we construct three evaluation benchmarks, that include a large variety of common script generation scenarios. Experiments show our framework achieves a significant improvement on diverse script generation tasks, highlighting its potential for specialized custom tools.

\clearpage

\bibliographystyle{abbrv}
\bibliography{references}

\begin{thebibliography}{10}

\bibitem{CodeLlama}
M.~AI.
\newblock Codellama: A large language model for code generation, 2024.

\bibitem{StableCode}
S.~AI.
\newblock Stablecode: A large language model for code generation, 2024.

\bibitem{Codex}
M.~Chen, J.~Tworek, H.~Jun, Q.~Yuan, H.~Pond{\'e}, et~al.
\newblock Evaluating large language models trained on code.
\newblock {\em ArXiv/2107.03374}, 2021.

\bibitem{fakhoury2024ltestdriven}
S.~Fakhoury, A.~Naik, G.~Sakkas, S.~Chakraborty, and S.~K. Lahiri.
\newblock Llm-based test-driven interactive code generation: User study and empirical evaluation.
\newblock {\em arXiv/2404.10100}, 2024.

\bibitem{NEURIPS2023_AGI}
Y.~Ge, W.~Hua, K.~Mei, j.~ji, J.~Tan, S.~Xu, Z.~Li, and Y.~Zhang.
\newblock Openagi: When llm meets domain experts.
\newblock In A.~Oh, T.~Naumann, A.~Globerson, K.~Saenko, M.~Hardt, and S.~Levine, editors, {\em Advances in Neural Information Processing Systems}, pages 5539--5568, 2023.

\bibitem{gong2024astt5}
L.~Gong, M.~Elhoushi, and A.~Cheung.
\newblock Ast-t5: Structure-aware pretraining for code generation and understanding.
\newblock {\em arXiv/2401.03003}, 2024.

\bibitem{golang}
Q.~Gu.
\newblock Llm-based code generation method for golang compiler testing.
\newblock In {\em Proceedings of the 31st ACM Joint European Software Engineering Conference and Symposium on the Foundations of Software Engineering}, page 2201–2203, 2023.

\bibitem{guo2024SDG}
X.~Guo and Y.~Chen.
\newblock Generative ai for synthetic data generation: Methods, challenges and the future.
\newblock {\em arXiv/2403.04190}, 2024.

\bibitem{jain2023CodeCleaning}
N.~Jain, T.~Zhang, W.-L. Chiang, J.~E. Gonzalez, K.~Sen, and I.~Stoica.
\newblock Llm-assisted code cleaning for training accurate code generators.
\newblock {\em arXiv/2311.14904}, 2023.

\bibitem{jain2024llmagents}
S.~Jain, A.~Dora, K.~S. Sam, and P.~Singh.
\newblock Llm agents improve semantic code search.
\newblock {\em arXiv/2408.11058}, 2024.

\bibitem{RAG_based_codegen}
H.~Koziolek, S.~Gr\"{u}ner, R.~Hark, V.~Ashiwal, S.~Linsbauer, and N.~Eskandani.
\newblock Llm-based and retrieval-augmented control code generation.
\newblock In {\em Proceedings of the 1st International Workshop on Large Language Models for Code}, page 22–29, 2024.

\bibitem{lin2024SoftwareUse}
F.~Lin, D.~J. Kim, Tse-Husn, and Chen.
\newblock When llm-based code generation meets the software development process.
\newblock {\em arXiv/2403.15852}, 2024.

\bibitem{Ansys2024EngCode}
Y.-C. Lin, A.~Kumar, N.~Chang, W.~Zhang, M.~Zakir, R.~Apte, H.~He, C.~Wang, and J.-S.~R. Jang.
\newblock Novel preprocessing technique for data embedding in engineering code generation using large language model.
\newblock {\em arXiv/2311.16267}, 2024.

\bibitem{chipnemo}
M.~Liu, T.-D. Ene, R.~Kirby, C.~Cheng, N.~Pinckney, R.~Liang, J.~Alben, H.~Anand, K.~Kunal, Ismet, et~al.
\newblock Chipnemo: Domain-adapted llms for chip design.
\newblock {\em arXiv/2311.00176}, 2023.

\bibitem{liu2024chatqa}
Z.~Liu, W.~Ping, R.~Roy, P.~Xu, C.~Lee, M.~Shoeybi, and B.~Catanzaro.
\newblock Chatqa: Surpassing gpt-4 on conversational qa and rag.
\newblock {\em arXiv/2401.10225}, 2024.

\bibitem{long2024SDG}
L.~Long, R.~Wang, R.~Xiao, J.~Zhao, X.~Ding, G.~Chen, and H.~Wang.
\newblock On llms-driven synthetic data generation, curation, and evaluation: A survey.
\newblock {\em arXiv/2406.15126}, 2024.

\bibitem{code_understanding}
D.~Nam, A.~Macvean, V.~Hellendoorn, B.~Vasilescu, and B.~Myers.
\newblock Using an llm to help with code understanding.
\newblock In {\em Proceedings of the IEEE/ACM 46th International Conference on Software Engineering}, 2024.

\bibitem{execution}
A.~Ni, S.~Iyer, D.~Radev, V.~Stoyanov, W.-T. Yih, S.~Wang, and X.~V. Lin.
\newblock {LEVER}: Learning to verify language-to-code generation with execution.
\newblock In {\em Proceedings of the 40th International Conference on Machine Learning}, volume 202, pages 26106--26128, 2023.

\bibitem{nvidia2024nemotron4340b}
Nvidia, :, B.~Adler, N.~Agarwal, A.~Aithal, D.~H. Anh, P.~Bhattacharya, A.~Brundyn, J.~Casper, et~al.
\newblock Nemotron-4 340b technical report.
\newblock {\em arXiv/2406.11704}, 2024.

\bibitem{StarCoder}
B.~Project.
\newblock Starcoder: A large language model for code generation, 2024.

\bibitem{Qin2023ToolLLM}
Y.~Qin, S.~Liang, Y.~Ye, K.~Zhu, L.~Yan, Y.-T. Lu, Y.~Lin, X.~Cong, X.~Tang, B.~Qian, S.~Zhao, R.~Tian, R.~Xie, J.~Zhou, M.~H. Gerstein, D.~Li, Z.~Liu, and M.~Sun.
\newblock Toolllm: Facilitating large language models to master 16000+ real-world apis.
\newblock {\em arXiv 2307.16789}, 2023.

\bibitem{rafailov2024DPO}
R.~Rafailov, A.~Sharma, E.~Mitchell, S.~Ermon, C.~D. Manning, and C.~Finn.
\newblock Direct preference optimization: Your language model is secretly a reward model.
\newblock {\em arXiv/2305.18290}, 2024.

\bibitem{wei2023CoT}
J.~Wei, X.~Wang, D.~Schuurmans, M.~Bosma, B.~Ichter, F.~Xia, E.~Chi, Q.~Le, and D.~Zhou.
\newblock Chain-of-thought prompting elicits reasoning in large language models, 2023.

\bibitem{yao2023react}
S.~Yao, D.~Yu, J.~Zhao, I.~Shafran, K.~Narasimhan, and Y.~Cao.
\newblock React: Synergizing reasoning and acting in language models.
\newblock In {\em International Conference on Learning Representations (ICLR)}, 2023.

\bibitem{zhang2024codeagent}
K.~Zhang, J.~Li, G.~Li, X.~Shi, and Z.~Jin.
\newblock Codeagent: Enhancing code generation with tool-integrated agent systems for real-world repo-level coding challenges.
\newblock {\em arXiv/2401.07339}, 2024.

\end{thebibliography}

\end{document}